\documentclass[aps,prl,showpacs]{revtex4}

\usepackage{amssymb,amsmath}
\usepackage{graphicx}
\usepackage{bm}

\begin{document}

\title{Collective Phase Sensitivity}

\author{Yoji Kawamura$^1$, Hiroya Nakao$^{2,3}$, Kensuke Arai$^2$,
  Hiroshi Kori$^4$, and Yoshiki Kuramoto$^{5,6}$}

\affiliation{
  $^1$The Earth Simulator Center, Japan Agency for Marine-Earth Science and Technology, Yokohama 236-0001, Japan \\
  $^2$Department of Physics, Graduate School of Sciences, Kyoto University, Kyoto 606-8502, Japan \\
  $^3$Abteilung Physikalische Chemie, Fritz-Haber-Institut der
  Max-Planck-Gesellschaft, Faradayweg 4-6,
  14195 Berlin, Germany \\
  $^4$Division of Advanced Sciences, Ochadai Academic Production,
  Ochanomizu University,
  Tokyo 112-8610, Japan \\
  $^5$Research Institute for Mathematical Sciences, Kyoto University,
  Kyoto 606-8502, Japan \\
  $^6$Institute for Integrated Cell-Material Sciences, Kyoto
  University, Kyoto 606-8501, Japan }

\date{November 29, 2007 ; revised \today}


\begin{abstract}
  The collective phase response to a macroscopic external perturbation
  of a population of interacting nonlinear elements exhibiting
  collective oscillations is formulated for the case of
  globally-coupled oscillators.
  The macroscopic phase sensitivity is derived from the microscopic
  phase sensitivity of the constituent oscillators by a two-step phase
  reduction.
  We apply this result to quantify the stability of the macroscopic
  common-noise induced synchronization of two uncoupled populations of
  oscillators undergoing coherent collective oscillations.
\end{abstract}

\maketitle

Nature possesses various examples of systems composed of many
nonlinear elements in which, through mutual interactions, coherent
collective dynamics
emerges~\cite{Ref:Winfree01,Ref:Kuramoto84,Ref:Pikovsky01,Ref:Manrubia04,
  Ref:Izhikevich07,Ref:Brunel99,Ref:Strogatz00,Ref:Acebron05,
  Ref:Okuda91,Ref:Montbrio04,Ref:Hudson05,Ref:Kiss,Ref:Ukai07}.
The collective dynamics of the population is generally not a simple
superposition of the microscopic dynamics of the constituent elements;
due to nonlinearities, it can differ qualitatively.
Understanding how the microscopic properties give rise to the
macroscopic collective properties of the system, and formulating the
theory in a simple closed form at the macroscopic level is a
fundamental problem in nonlinear dynamics.

If the population undergoes a stable macroscopic limit-cycle
oscillation, it can most simply be described by a single macroscopic
phase variable.
By knowing the {\em collective phase response} of the macroscopic
oscillation with respect to external perturbations, we can describe
the dynamics of the population by the phase picture, much like for single
microscopic
oscillators~\cite{Ref:Winfree01,Ref:Kuramoto84,Ref:Pikovsky01,Ref:Manrubia04,
  Ref:Izhikevich07}.
The collective phase response can be measured operatively by
macroscopically applying an external perturbation to the population
without recourse to the microscopic
details~\cite{Ref:Winfree01,Ref:Hudson05,Ref:Ukai07}.
However, to understand how the microscopic dynamics of the constituent
elements and their mutual interactions conspire to become a
macroscopic phase response, it is necessarily to develop a statistical
mechanical approach.

In this paper, we formulate this problem for globally-coupled
oscillators as a simple model of macroscopic populations exhibiting
collective oscillations.
Global coupling is the simplest form of mutual interactions, which is
both conceptually and practically important because it serves as a
starting point for theoretical analysis of various network-coupled
dynamical systems and because it is realized in experimental systems
~\cite{Ref:Winfree01,Ref:Kuramoto84,Ref:Pikovsky01,Ref:Manrubia04,Ref:Acebron05,Ref:Kiss}.
We quantitatively derive the {\em collective phase sensitivity}, which
is the linear response coefficient of the macroscopic collective phase
response, from the microscopic phase sensitivity of the constituent
oscillators using a two-step phase-reduction on the micro- and
macroscopic scales.

The collective phase sensitivity is a fundamental quantity
characterizing the response of a macroscopic limit-cycle oscillation
to weak external perturbations.
As one application, we analyze macroscopic common-noise induced
synchronization of the collective oscillations between two uncoupled
populations.
Though not treated in this paper, the entrainment of a population
to external forcing or synchronization between populations can also
easily be analyzed using the collective phase sensitivity.

\begin{figure}[htbp]
  \begin{center}
   \includegraphics[width=0.65\hsize,clip]{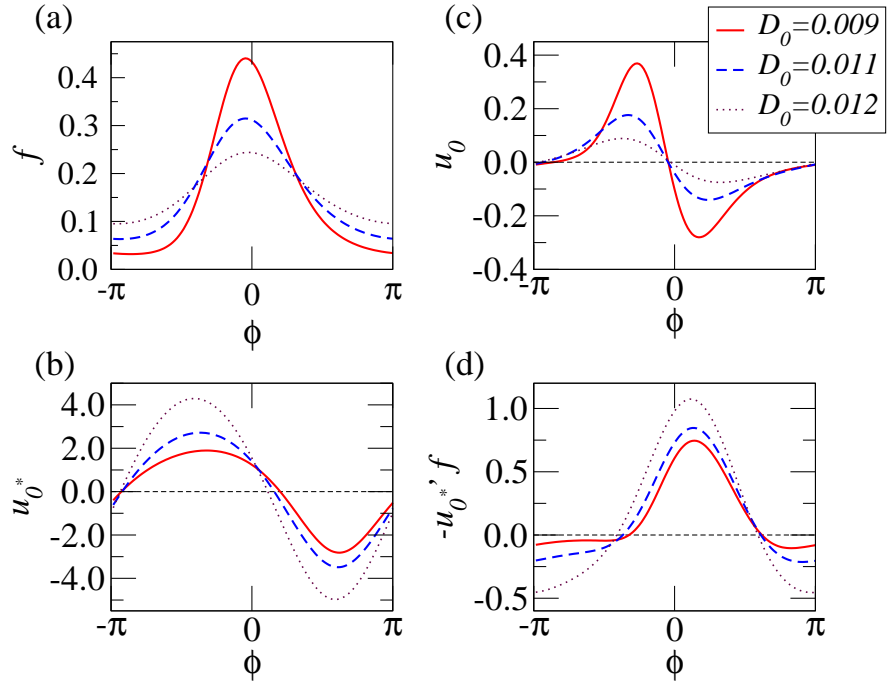}
    \caption{
      (color online).
      Globally-coupled SL model.
      (a) Stationary PDF $f(\phi)$, (b) zero eigenfunction
      $u_0^\ast(\phi)$ of $\hat{L}^\ast$, (c) zero eigenfunction
      $u_0(\phi)$ of $\hat{L}$, and (d) kernel $-u_0^\ast(\phi)'
      f(\phi)$.  }
    \label{Fig:1}
  \end{center}
\end{figure}

\begin{figure}[htbp]
  \begin{center}
    \includegraphics[width=0.8\hsize,clip]{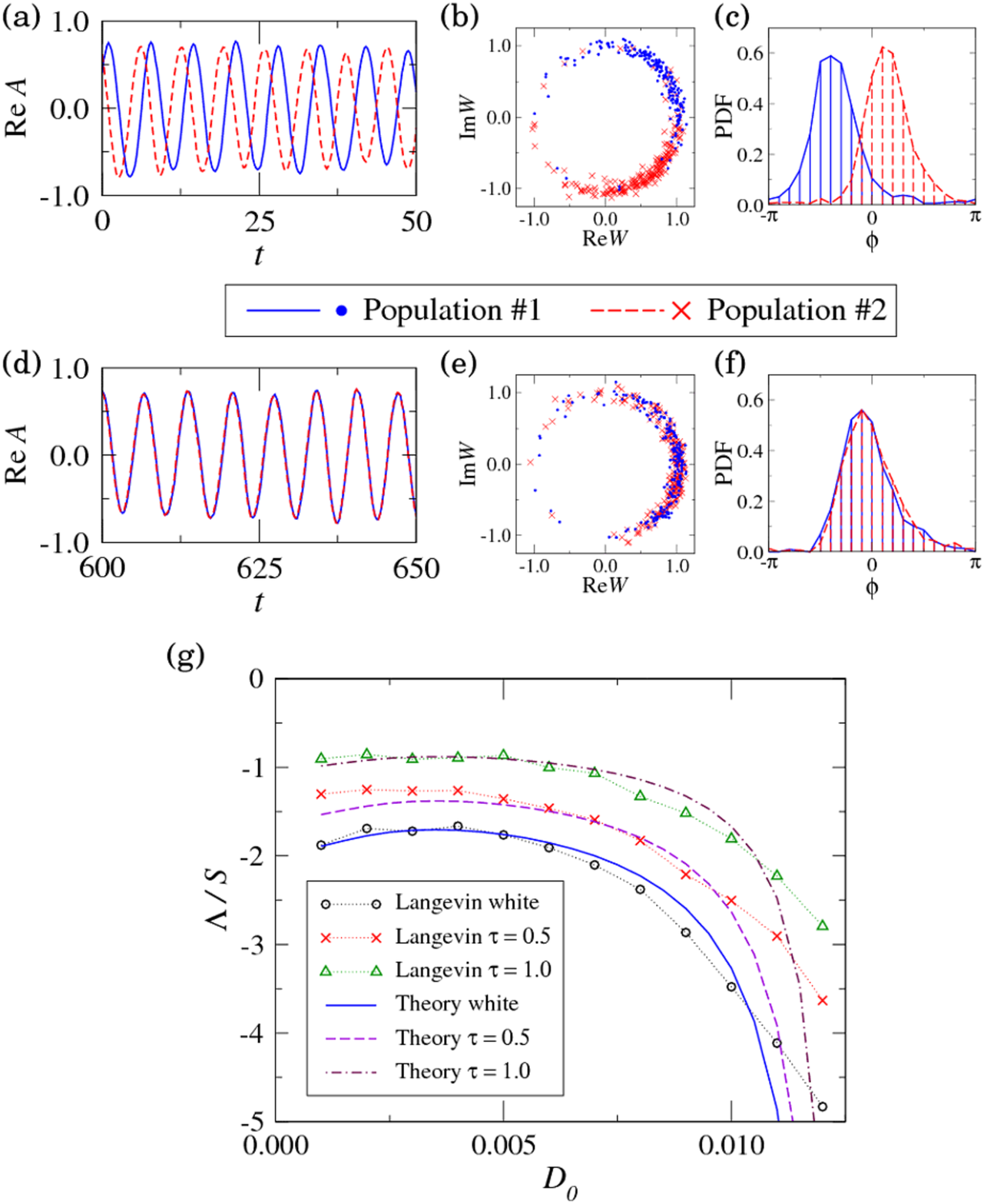}
    \caption{
      (color online).
      Common-noise induced synchronization of globally-coupled SL populations.
      Time evolution of the real part of $A(t)$ from (a) $t = 0$ and (d) $t = 600$.
      Snapshots of $\{W_j\}$ at (b) $t = 0$ and (e) $t = 600$
      (1/5 of population shown).
      PDF of phase at (c) $t = 0$ and (f) $t = 600$.
      (g) Comparison between Lyapunov exponent $\Lambda$, found from
      theory and Langevin simulation.  }
    \label{Fig:3}
  \end{center}
\end{figure}

Let us give a general definition of the collective phase sensitivity
first.
We consider a population of globally-coupled noisy identical
limit-cycle oscillators collectively exhibiting stable macroscopic
limit-cycle oscillations.
We assume that the effect of coupling, noise, and perturbations is
sufficiently weak, so that the orbit of each microscopic oscillator is
always near its limit cycle.
We can then describe each oscillator by the phase $\phi$, and the
whole population by a single-oscillator phase probability density
function (PDF) $P(\phi, t)$ of the constituent oscillators.
As the population undergoes limit-cycle oscillations, $P(\phi, t)$ is
given in the form of a rotating wave packet of constant shape (see
Figs.~\ref{Fig:1} and~\ref{Fig:3}),
\begin{align}
  P(\phi, t) = f\left(\phi - \Phi(t)\right),
  \;\;\;
  \Phi(t) = \Omega t + \Theta,
  \label{Eq:PDF1}
\end{align}
where $f(\phi)$ represents the wave packet, $\Phi(t)$ the collective
phase of the population (location of the wave packet) at time $t$,
$\Omega$ the frequency of the collective oscillation, and $\Theta$ the
initial collective phase.
We assume $\phi \in [-\pi, \pi]$ and impose a periodic boundary
condition $f(\phi+2\pi) = f(\phi)$.

At the instant that the collective phase is $\Phi_0$, we apply a
macroscopic perturbation of magnitude and direction ${\bm s}$ that
acts uniformly on all oscillators.
The shape of $P(\phi, t)$ transiently deforms, but due to the
stability of the collective oscillations, eventually settles into its
stationary shape, Eq.~(\ref{Eq:PDF1}).
We denote the perturbed PDF after a sufficiently long time $t_0$ as
$P_1(\phi, t+t_0) = f\left(\phi - \Phi_1(\Phi_0, {\bm s}, t_0)\right)$,
and the unperturbed PDF after time $t_0$ as
$P_2(\phi, t+t_0) = f\left(\phi - \Phi_2(\Phi_0, t_0)\right)$,
where $\Phi_2 = \Phi_0 + \Omega t_0$.
The collective phase response function $G(\Phi_0; {\bm
    s})$ is defined as the asymptotic difference of $\Phi_1(\Phi_0,
{\bm s}, t_0)$ and $\Phi_2(\Phi_0, t_0)$: $ G(\Phi_0; {\bm s}) =
\lim_{t_0 \to \infty} \left[ \Phi_1(\Phi_0, {\bm s}, t_0) -
  \Phi_2(\Phi_0, t_0) \right] $~\cite{Ref:Winfree01}.
This is in general a function of $\Phi_0$ and ${\bm s}$.
When the impulse magnitude $|{\bm s}|$ is small enough, the collective
phase response function is proportional to ${\bm s}$, and is given by
$ G(\Phi_0; {\bm s}) = \boldsymbol{\zeta}(\Phi_0) \cdot {\bm s} $,
where $\boldsymbol{\zeta}(\Phi_0)$ is the collective phase
sensitivity~\cite{Ref:Kuramoto84}.

We consider globally-coupled oscillators described by the following
Langevin equation:
\begin{align}
  \dot{\bm X}_j(t) = {\bm F}( {\bm X}_j(t) )
  + \frac{1}{N} \sum_{k=1}^{N} {\bm V}( {\bm X}_k(t) )
  + \sqrt{2 D_0}\, {\boldsymbol \eta}_j(t)
  + {\bm s}(t),
  \label{Eq:GCO}
\end{align}
for $j = 1, \cdots, N$, where ${\bm X}_j$ is the state of the $j$th
oscillator, ${\bm F}({\bm X})$ the dynamics of a single oscillator,
${\bm V}({\bm X})$ the contribution from each oscillator to the mean
field, $D_0$ the noise intensity, ${\boldsymbol \eta}_j(t)$ an
independent zero-mean Gaussian white noise of unit intensity, which
satisfies $\langle {\boldsymbol \eta}_j(t) {\boldsymbol \eta}_k(t')
\rangle = {\bf I} \delta_{j,k} \delta(t - t')$ where ${\bf I}$ is the
unit matrix, and ${\bm s}(t)$ the macroscopic external perturbation
common to all oscillators.
When isolated, each oscillator has a stable limit-cycle ${\bm
  X}_0(t+T) = {\bm X}_0(t)$ with period $T = 2\pi / \omega$.
We define a phase $\phi({\bm X}) \in [-\pi, \pi]$ around the limit
cycle, which increases with a constant natural frequency
$\omega$~\cite{Ref:Kuramoto84}.
We assume that the effect of coupling, noise, and perturbation is
sufficiently weak.

To derive a dynamical equation for the single-oscillator phase PDF of
our system, we perform the first, microscopic phase reduction of
Eq.~(\ref{Eq:GCO}).
Using the microscopic phase sensitivity of each oscillator, ${\bm
  Z}(\phi) = \nabla_{\bm X} \phi |_{ {\bm X} = {\bm
    X}_0(\phi)}$~\cite{Ref:Kuramoto84}, the dynamics of the system can
be reduced to the following Langevin phase equation:
\begin{align}
  \dot{\phi}_j(t) = \omega + \frac{1}{N} \sum_{k=1}^{N} {\bm
    Z}(\phi_j) \cdot {\bm V}({\bm X}_0(\phi_k)) + \sqrt{2 D_0}\, {\bm
    Z}(\phi_j) \cdot {\boldsymbol \eta}_j(t) + \xi(\phi_j, t),
\end{align}
where the effect of the perturbation is given by $\xi(\phi_j, t) =
{\bm Z}(\phi_j) \cdot {\bm s}(t)$.
From this equation, after averaging and taking the limit as $N \to
\infty$ (see~\cite{Ref:Kuramoto84,Ref:Kawamura07} for details), we
obtain the nonlinear Fokker-Planck equation (FPE) for the
single-oscillator phase PDF $P(\phi, t)$,
\begin{align}
  \frac{\partial}{\partial t} P(\phi, t)
  &= -\frac{\partial}{\partial \phi} 
  \left\{ \left[
    \omega + \int_{-\pi}^{\pi} \Gamma( \phi - \phi' ) P(\phi', t) d\phi' + \xi(\phi, t)
    \right] P(\phi, t) \right\}
  + D \frac{\partial^2}{\partial \phi^2} P(\phi, t),
  \label{Eq:FPE0}
\end{align}
where the phase coupling function is given by
$\Gamma( \phi - \phi' ) = (1 / 2\pi)
\int_{-\pi}^{\pi} {\bm Z}(\phi + \psi) \cdot {\bm V}(\phi' + \psi) d\psi$,
and the diffusion coefficient by
$D = (D_0 / 2 \pi) \int_{-\pi}^{\pi} {\bm Z}(\psi) \cdot {\bm Z}(\psi)
d\psi$.
The external perturbation $\xi(\phi, t)$ in the drift term is not
averaged, but kept as a time-dependent term.

To realize stable macroscopic oscillations, we assume that the
coupling function $\Gamma( \phi )$ is unimodal and satisfies the
in-phase (attractive) condition $d\Gamma( \phi ) / d\phi|_{\phi = 0} <
0$~\cite{Ref:Kuramoto84}.
When there is no external perturbation ($\xi(\phi, t)\equiv 0$), the
solution of Eq.~(\ref{Eq:FPE0}) is given by a rotating wave packet of
the form Eq.~(\ref{Eq:PDF1}), which behaves as follows:
In the absence of independent noise ($D_0 = 0$), the oscillators are
completely synchronized, so that $P(\phi, t)$ is delta-peaked.
As $D_0$ is increased, $P(\phi, t)$ takes on a finite width while
rotating at a constant speed.
As $D_0$ increases past a critical value $D_{\rm c}$, the collective
oscillation disappears, and $P(\phi, t) \equiv 1/2\pi$.
For the following analysis, we take $0 < D_0 < D_{\rm c}$.

We now derive the collective phase sensitivity
$\boldsymbol{\zeta}(\Phi)$ from the microscopic phase sensitivity
${\bm Z}(\phi)$ by seeking a closed equation for the collective phase
$\Phi(t)$.
This is performed by the second, macroscopic phase reduction,
similarly to the derivation of wavefront dynamics for oscillatory
media~\cite{Ref:Kuramoto84}.  Our treatment here is a generalization
of~\cite{Ref:Kawamura07} for nonlocally-coupled noisy oscillators. 
Let us ignore the perturbation for the moment ($\xi(\phi, t) \equiv
0$).
Introducing a corotating phase with the wave packet, $\theta = \phi -
\Phi(t) = \phi - \Omega t - \Theta$, the solution of
Eq.~(\ref{Eq:FPE0}), $P(\phi, t) = f(\theta)$, satisfies $ D ( d^2 / d
\theta^2 ) f(\theta) - ( d / d \theta ) \left\{ \left[ (\omega -
    \Omega) + \int_{-\pi}^{\pi} \Gamma(\theta - \theta') f(\theta')
    d\theta' \right] f(\theta) \right\} = 0 $.
Small deviations $u(\theta)$ from this unperturbed wave packet
$f(\theta)$ obey a linearized equation $ \partial_t u(\theta, t) =
\hat{L} u(\theta, t) $, where the linear operator $\hat{L}$ is given
by
\begin{align}
  \hat{L} u(\theta) =
  D \frac{d^2}{d \theta^2} u(\theta) - \frac{d}{d
    \theta} \left\{ \left[ (\omega - \Omega) + \int_{-\pi}^{\pi}
      \Gamma(\theta - \theta') f(\theta') d\theta' \right] u(\theta)
  \right\}
%%  \cr
  - \frac{d}{d \theta} \left\{ f(\theta) \int_{-\pi}^{\pi}
    \Gamma(\theta - \theta') u(\theta') d\theta' \right\}.
\end{align}
Defining the inner product as $ [ u^\ast (\theta), u(\theta) ] =
\int_{-\pi}^{\pi} u^\ast (\theta) u(\theta) d\theta $, we introduce an
adjoint operator $\hat{L}^\ast$ of $\hat{L}$ by $ [ u^\ast(\theta),
\hat{L} u(\theta) ] = [ \hat{L}^\ast u^\ast(\theta), u(\theta) ] $.
In the calculation below, we only need zero eigenfunctions
$u_0(\theta)$ of $\hat{L}$ and $u_0^{\ast}(\theta)$ of
$\hat{L}^{\ast}$, which we normalize as $[u^\ast_{0}(\theta),
u_{0}(\theta)] = 1$.
It is easy to check that $u_0(\theta)$ can be chosen as $ u_0(\theta)
= d f(\theta) / d \theta $, reflecting the translational symmetry of
Eq.~(\ref{Eq:FPE0})~\cite{Ref:Kuramoto84,Ref:Kawamura07}.

We now incorporate the effect of perturbation, $\xi(\phi, t)$.
To obtain a closed equation for $\Phi(t)$, we assume that the solution
of Eq.~(\ref{Eq:FPE0}) is still given in the form of
Eq.~(\ref{Eq:PDF1}), ignoring its slight deformation at the lowest
order.  Instead, we allow the constant $\Theta$ in Eq.~(\ref{Eq:PDF1})
to vary slowly with time as $\Theta(t)$.
Plugging $P(\phi, t) = f(\theta(t)) = f(\phi - \Omega
  t - \Theta(t))$ into Eq.~(\ref{Eq:FPE0}), we obtain $ u_0(\theta)
\dot{\Theta}(t) = (d / d \theta) \left\{ \xi(\theta + \Omega t +
  \Theta(t), t) f(\theta) \right\} $.
Taking the inner product with $u_0^{\ast}(\theta)$ on both sides
yields
%%
%%\begin{align}
$ \dot{\Theta}(t) = \int_{-\pi}^{\pi} u_0^{\ast}(\theta) (d / d
\theta) \left\{ \xi(\theta + \Omega t + \Theta(t), t) f(\theta)
\right\} d\theta.  $
%%\end{align}
%%
Noting that $\xi(\theta + \Omega t + \Theta(t), t) = \xi(\theta +
\Phi(t), t) = {\bm Z}(\theta + \Phi(t)) \cdot {\bm s}(t)$, we arrive
at the {\em macroscopic phase equation} obeyed by $\Phi(t)$,
\begin{align}
  \dot{\Phi}(t) = \Omega + {\boldsymbol \zeta}(\Phi(t)) \cdot {\bm s}(t),
  \label{Eq:colphaseeq}
\end{align}
where the {\em collective phase sensitivity} is given by
\begin{align}
  {\boldsymbol \zeta}(\Phi) 
  &= \int_{-\pi}^{\pi} \left( - \frac{d u_0^\ast(\theta)}{d \theta} f(\theta) \right)
  {\bm Z}(\theta + \Phi) d\theta.
  \label{Eq:colZ}
\end{align}
Thus, ${\boldsymbol \zeta}(\Phi)$ is expressed as a convolution of the
microscopic phase sensitivity ${\bm Z}(\theta)$ with a kernel $-
u_0^\ast(\theta)' f(\theta)$.
Note the kernel is not merely the PDF $f(\theta)$ as we would naively
expect.

Now let us illustrate our results with an example of $N$ noisy
globally-coupled Stuart-Landau (SL) oscillators described by the
following Langevin equation~\cite{Ref:Kuramoto84,Ref:Kawamura07}:
$\dot{W}_j(t) = (1 + i \omega_0) W_j - (1 + i \beta) |W_j|^2 W_j
+ ( K / N ) \sum_{k=1}^{N} W_k + \sqrt{2 D_0}\, \eta_j(t) + s(t)$.
Here, the complex amplitude $W_j = W^{\rm R}_j + i W^{\rm I}_j$ (${\rm
  R}$ and ${\rm I}$ express real and imaginary components,
respectively) describes the state of the $j$th oscillator (i.e. ${\bm
  X}_j = (W^{\rm R}_j, W^{\rm I}_j)$), $\omega_0$ and $\beta$ are
oscillator parameters, $K$ the coupling strength, $s(t) = s^{\rm R}(t)
+ i s^{\rm I}(t)$ the macroscopic complex perturbation, $D_0$ the
noise intensity, and $\eta_j(t) = \eta^{\rm R}_j(t) + i \eta^{\rm
  I}_j(t)$ the independent complex Gaussian white noise of zero mean
and unit intensity which satisfies $\langle \eta^{\rm R}_j(t)
\eta^{\rm R}_k(t') \rangle = \langle \eta^{\rm I}_j(t) \eta^{\rm
  I}_k(t') \rangle = \delta_{j,k} \delta(t-t')$ and $\langle \eta^{\rm
  R}_j(t) \eta^{\rm I}_k(t') \rangle = 0$.
The phase $\phi$ is defined on the complex plane as $\phi = \arg W -
\beta \ln |W|$, which grows constantly with a natural frequency
$\omega = \omega_0 - \beta$ in the absence of coupling and other
external effects.
The microscopic phase sensitivity has both real and imaginary parts,
given analytically as
${\bm Z}(\phi) = ( Z^{\rm R}(\phi), Z^{\rm I}(\phi) ) = ( - \sin \phi
- \beta \cos \phi,\, \cos \phi - \beta \sin \phi )$,
and the phase coupling function is given by
$\Gamma( \phi - \phi' ) = - K \sqrt{ 1 + \beta^2 } \sin( \phi - \phi'
+ \alpha ), \, \alpha = \arg( 1 + i \beta
)$~\cite{Ref:Kuramoto84,Ref:Kawamura07}.
We use the complex order parameter $A(t) = R(t) e^{ i \tilde{\Phi}(t)
} = (1 / N) \sum_{k=1}^{N} e^{ i \phi_k(t) } \simeq \int_{-\pi}^{\pi}
e^{ i \phi } f( \phi - \Phi(t) ) d\phi$ to quantify the collective
oscillation, whose modulus $R(t)$ serves as a measure of the
coherence~\cite{Ref:Kuramoto84,Ref:Kawamura07}.
By choosing $f(\phi)$ so that $\int_{-\pi}^{\pi} e^{i \phi} f(\phi)
d\phi$ is positive and real, and taking the initial collective phase
as $\Theta = 0$, namely, $P(\phi, 0) \equiv f(\phi)$, the phase
$\tilde{\Phi}(t)$ of the order parameter $A(t)$ matches the
collective phase $\Phi(t)$ defined in Eq.~(\ref{Eq:PDF1}).

\begin{figure}[htbp]
  \begin{center}
    \includegraphics[width=0.65\hsize,clip]{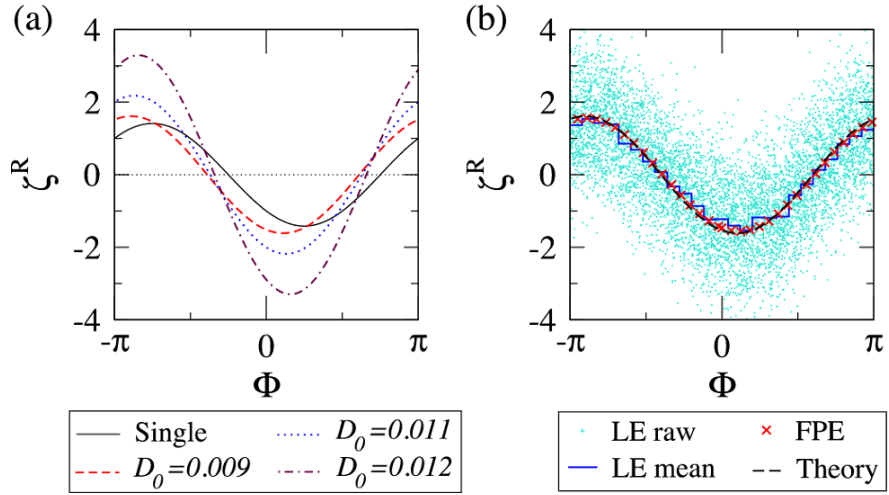}
    \caption{
      (color online).
      Collective phase sensitivity of globally-coupled SL model. 
      (a) Theoretical curves of $\zeta^{\rm R}(\Phi)$.
      Single-oscillator phase sensitivity $Z^{\rm R}(\phi)$ is also
      shown with $\phi = \Phi$.
      (b) $\zeta^{\rm R}(\Phi)$, found from theory and from
      numerical simulations of nonlinear FPE and Langevin equation (LE).
      For the LE simulation, individual raw responses are shown as dots
      and piecewise mean values (25 bins) are shown as a solid line.
    }
    \label{Fig:2}
  \end{center}
\end{figure}

Figure~\ref{Fig:1} shows the stationary PDF $f(\phi)$, the zero
eigenfunctions $u_0(\phi)$ and $u_0^\ast(\phi)$, and the kernel
$-u_0^{\ast}(\phi)' f(\phi)$, obtained by numerically solving
Eq.~(\ref{Eq:FPE0}).
The parameters $\omega_0 = 2$, $\beta = 1$, $K=0.05$ are fixed, while
the noise intensity is varied as $D_0 = 0.009$, $0.011$, $0.012$.
As $D_0$ approaches the critical value $D_{\rm c} = 0.0125$
~\cite{Ref:Kuramoto84,Ref:Kawamura07}, $f(\phi)$ becomes flat, while
the amplitudes of $u_0^{\ast}(\phi)$ and the kernel $-u_0^{\ast}(\phi)'
f(\phi)$ increase.
Figure~\ref{Fig:2}(a) shows the real part of the collective phase
sensitivity $\zeta^{\rm R}(\Phi)$ calculated using the results shown
in Fig.~\ref{Fig:1} (the imaginary part is simply given by $\zeta^{\rm
  I}(\Phi) = \zeta^{\rm R}(\Phi - \pi/2)$ due to symmetry of the SL
oscillator).
For comparison, we show the microscopic phase sensitivity, $Z^{\rm
  R}(\phi)$ with $\phi = \Phi$, which corresponds to the $D_0 \to 0$
limit.
For $D_0 > 0$, $\zeta^{\rm R}(\Phi)$ is different from $Z^{\rm
  R}(\phi)$ due to distributed individual phases.
As $D_0 \to D_{\rm c}$, $\zeta^{\rm R}(\Phi)$ diverges as $(D_{\rm c}
- D_0)^{-1/2}$~\cite{footnote2}.
Figure~\ref{Fig:2}(b) compares the theoretical $\zeta^{\rm R}(\Phi)$
at $D_0 = 0.009$ with those directly measured by adding sufficiently
weak impulses ($|s| = 0.001$) to the nonlinear FPE~(\ref{Eq:FPE0}) and
to the Langevin equation~(\ref{Eq:GCO}) with $N = 10,000$
oscillators~\cite{footnote3}.
The results of the nonlinear FPE corresponding to the limit $N \to
\infty$ agrees well with the theory~\cite{footnote4}.
The results of the Langevin simulation show a wide distribution of the
values because $N$ is necessarily finite, though the piecewise average
value shows reasonable agreement with the theory~\cite{footnote5}.

As an application of the collective phase sensitivity, we analyze the
stability of the synchronized state of two uncoupled populations of
collectively oscillating, globally-coupled SL oscillators due to a
macroscopic common noise.
Consider two uncoupled macroscopic phase oscillators driven by a
common, weak Gaussian noise ${\boldsymbol s}(t)$:
$\dot{\Phi}^{(\sigma)}(t) = \Omega + {\boldsymbol
  \zeta}(\Phi^{(\sigma)}(t)) \cdot {\boldsymbol s}(t) \ (\sigma=1,
2)$, where the noise correlation is given by $\langle
\boldsymbol{s}(t) \boldsymbol{s}(0) \rangle = {\bf I} C(t)$.
In~\cite{Ref:Teramae04}, it is shown that a weak common Gaussian noise
will cause synchronization of uncoupled oscillators, and the Lyapunov
exponent that quantifies the growth rate of an infinitesimal phase
difference is given by
$\Lambda = (1 / 2 \pi) \int_{0}^{\infty} dt \, C(t)
\int_{-\pi}^{\pi} d\Phi \, {\boldsymbol \zeta}''(\Phi)
\cdot {\boldsymbol \zeta}(\Phi - \Omega t)$,
where $\Lambda \leq 0$ always holds.
For the globally-coupled SL oscillators, the perturbation
${\boldsymbol s}(t)$ has real and imaginary components.
We use the Ornstein-Uhlenbeck process~\cite{Ref:Gardiner97},
$\dot{z}(t) = -z/\tau + \chi(t)/\tau$ with $\chi(t)$ a zero-mean
Gaussian white noise of unit intensity, to create colored Gaussian
noises $z^{\rm R, I}(t)$ with the correlation time $\tau$, and put
${\boldsymbol s}(t) = \sqrt{2 S} (z^{\rm R}(t),\, z^{\rm I}(t))$ where
$S$ controls the intensity of common noise.  The correlation function
is given as $C(t) = S \exp( -|t|/\tau ) / \tau$.

Figure~\ref{Fig:3} shows the results for two uncoupled populations of
$N = 1000$ globally-coupled SL oscillators receiving macroscopic
common Gaussian noise of intensity $S = 0.001$ with correlation times
$\tau = 1.0$, $\tau = 0.5$, and $\tau \to 0$ (white limit).
Other parameters are $\omega_0 = 2$, $\beta = 1$, $K=0.05$, and $D_0 =
0.007$.
Figures~\ref{Fig:3}(a) and \ref{Fig:3}(d) show the time evolution of
the real part of the order parameter $A(t)$ of the two populations
obtained by the Langevin simulation.
At time $t = 0$, the two are not synchronized, but by $t = 600$, they
show near-complete synchronization, demonstrating that macroscopic
common noise can cause the collective oscillations of two uncoupled
populations to synchronize.
We stress that this is purely a macroscopic phenomenon of the
collective phase becoming entrained; the oscillators in the two
populations never become individually synchronized.
Figures~\ref{Fig:3}(b) and \ref{Fig:3}(e) show the distributions of
the states $\{W_j\}$ of the microscopic oscillators on a complex plane
at $t = 0$ and $t = 600$, and Figs.~\ref{Fig:3}(c) and \ref{Fig:3}(f)
show the corresponding histogram of the phases $\{ \phi_j \}$.
At $t = 0$, the distributions of oscillators only slightly overlap,
while by $t = 600$, they overlap considerably.
Figure~\ref{Fig:3}(g) compares the Lyapunov exponent $\Lambda$ as
measured from the Langevin simulation, averaged over $1500$ sample
paths, with the theoretical results.
Despite the relatively large fluctuations shown by the collective
oscillations, we see that $\Lambda$ shows good agreement with theory.
As $D_0$ approaches the critical value $D_{\rm c}$, the amplitude of
the collective phase sensitivity ${\boldsymbol \zeta}(\Phi)$
increases, so $\Lambda$ takes on ever larger, negative values.
However, near the critical point, the phase response becomes strongly
nonlinear, so that the numerical results diverge from the linear
theory based on the collective phase sensitivity.

In summary, we derived the macroscopic collective phase sensitivity
from the microscopic phase sensitivity
for a population of globally-coupled oscillators, and analyzed the
common-noise induced synchronization of two such populations.
By virtue of the assumption that the constituent oscillators are only
weakly perturbed, we could utilize the phase reduction method to
construct a general framework for the collective phase sensitivity,
which may bring a new perspective to the existing body of research on
coupled collective
oscillations~\cite{Ref:Okuda91,Ref:Montbrio04,Ref:Hudson05}.
Since we treated the SL oscillators as an example, the resulting
collective phase sensitivity was always sinusoidal, with only a change
in amplitude.
For other types of oscillators, the shape of the collective phase sensitivity
may differ significantly than that of its constituent oscillators.
Furthermore, the collective phase response to strong macroscopic
perturbations should prove to be even more intriguing, though our
present framework based on phase reduction is not applicable in this
scenario.
More detailed and generalized analysis will be reported in the near
future.

%%\begin{acknowledgments}
%%  This work was supported by the 21st Century COE program
%%  ``Center for Diversity and Universality in Physics''
%%  from MEXT Japan.
%%\end{acknowledgments}

\newpage

\vspace{1cm}

\begin{thebibliography}{99}
\bibitem{footnote2} This is because near a supercritical Hopf
  bifurcation, the amplitude of the deviation of $f(\phi)$ from the
  uniform PDF $1/2\pi$ scales as $( D_{\rm c} - D_0 )^{1/2}$, so that
  the amplitudes of $u_0^{\ast}(\phi)$ and the kernel
  scale accordingly as $( D_{\rm c} - D_0 )^{-1/2}$.
\bibitem{footnote4} Because the impulse amplitude $|s|$ is finite
  though small, there is a slight deviation in the result which
  becomes more pronounced as $D_0$ approaches $D_{\rm c}$.
\bibitem{footnote5} For the discussion of noise-induced
  synchronization, the mean, and not the fluctuation, is the vital
  factor.
\end{thebibliography}
\end{document}